\documentclass[runningheads]{llncs}
\usepackage{graphicx}
\usepackage{amsmath,amssymb,amsfonts}
\usepackage{algorithmic}
\usepackage{graphicx}
\usepackage{textcomp}
\usepackage{lineno,hyperref}
\usepackage{booktabs}
\usepackage{multirow}
\usepackage{multicol}
\usepackage{color}
\usepackage{bm}
\usepackage[misc,geometry]{ifsym}
\usepackage{hyperref}
\begin{document}
\title{SuperVessel: Segmenting High-resolution Vessel from Low-resolution Retinal Image}
\titlerunning{SuperVessel: Segmenting High-resolution Vessel}
%
%
\author{Yan Hu\inst{1}\thanks{Corresponding Author} \and Zhongxi Qiu\inst{1} \and Dan Zeng\inst{1} \and Li Jiang\inst{2} \and Chen Lin\inst{2} \and Jiang Liu\inst{1,3}}
\authorrunning{Yan Hu et al.}
\institute{Research Institute of Trustworthy Autonomous Systems and Department of Computer Science and Technology, Southern University of Science and Technology, Shenzhen, 518055, China \\
\email{huy3@sustech.edu.cn}
\and 
Department of Ophthalmology, Shenzhen People's Hospital (The Second Clinical Medical College, Jinan University; The First Affiliated Hospital, Southern University of Science and Technology) \and Guangdong Provincial Key Laboratory of Brain-inspired Intelligent Computation, Department of Computer Science and Engineering, Southern University of Science and Technology, Shenzhen, 518055, China
}
\maketitle              
\begin{abstract}
Vascular segmentation extracts blood vessels from images and serves as the basis for diagnosing various diseases, like ophthalmic diseases. Ophthalmologists often require high-resolution segmentation results for analysis, which leads to super-computational load by most existing methods. If based on low-resolution input, they easily ignore tiny vessels or cause discontinuity of segmented vessels. To solve these problems, the paper proposes an algorithm named SuperVessel, which gives out high-resolution and accurate vessel segmentation using low-resolution images as input. We first take super-resolution as our auxiliary branch to provide potential high-resolution detail features, which can be deleted in the test phase. Secondly, we propose two modules to enhance the features of the interested segmentation region, including an upsampling with feature decomposition (UFD) module and a feature interaction module (FIM) with a constraining loss to focus on the interested features. Extensive experiments on three publicly available datasets demonstrate that our proposed SuperVessel can segment more tiny vessels with higher segmentation accuracy IoU over 6\%, compared with other state-of-the-art algorithms. Besides, the stability of SuperVessel is also stronger than other algorithms. We will release the code after the paper is published.

\keywords{Vessel Segmentation  \and Super-resolution \and Multi-task Learning \and Retinal Image.}
\end{abstract}
\section{Introduction}
Retinal images are widely adopted as effective tools for the diagnosis and therapy of various diseases. The visual exploration of retinal blood vessels assists ophthalmologists in diagnosing variety of abnormalities of eyes, such as diabetic retinopathy, glaucoma, age-related macular degeneration. Researchers also proved the changes of retinal vessels could be an early screening method for some brain diseases \cite{london2013retina}, cardiovascular diseases \cite{ikram2013retinal}, or systematic diseases \cite{sun2009retinal}. Retinal vessel segmentation is one fundamental step for retinal image analysis. Identifying the vessel structures based on high-resolution images can give doctors great convenience of precise disease diagnosis. It brings a great burden and consumes plenty of time for doctors to segment vessels manually since it requires specific medical training and technical expertise. 

In recent years, researchers have proposed many automatic vessel segmentation algorithms based on deep-learning construction to lighten the burdens on doctors. They learn from the raw image data without adopting handcrafted features. Ronneberger et al. \cite{ronneberger2015u} proposed U-shape Net (U-Net) framework for biomedical image segmentation, which has become a popular neural network architecture for its promising results in biomedical image segmentation. Many variations have been proposed based on U-Net for different vessel segmentation tasks. For example, Fu et al. \cite{fu2016deepvessel} adopted the CRF to gather the multi-stage feature maps for improving the vessel detection performance. Some researchers proposed to stack multiple U-net shape architectures \cite{DBLP:conf/wacv/LiVNNK20}, input image patches into U-net architecture \cite{DBLP:conf/miccai/Wang0H19}, introduce multi-scale input layer to the conventional U-net \cite{su2021msu}, or cascade a backbone residual dense network and a fine-tune tail network \cite{karaali2021dr}. For computation loads, the existing algorithms often output low-resolution vessel segmentation results or direct upsampling results leading to a discontinuity in the results, which cannot satisfy the requirement of ophthalmologists. They often require high-resolution continuous vessels for analyzing diseases like branch retinal vein occlusion (BRVO), high-resolution (HR) images can provide more details as tiny vessels.

Recently, Wang et al. \cite{wang2021patch} proposed a patch-free 3D brain tumor segmentation driven by super-resolution technique. An HR 3D patch is necessary to guide segmentation and super-resolution during training, which may increase the computation complexity. In natural image segmentation, researchers proposed some auxiliary segmentation tasks \cite{DBLP:conf/igarss/LeiSWPXH19,Wang_2020_CVPR}, which adopt the feature loss between segmentation and super-resolution branches to indicate the task fusion. However, only constrained by image feature similarity cannot provide effective features for the vessel segmentation task, as vessel proportion in an entire image is relatively small. 

Therefore, to solve the above problems, we propose to output high-resolution vessel segmentation only based on low-resolution images, which supplies doctors with clear vessels for accurate diagnosis. Then we try to improve the vessel segment accuracy by focusing on our interested vessel regions with effective feature interaction. The contributions are as follows: 1) We propose a novel dual-stream learning algorithm that combines segmentation and super-resolution to produce the high-resolution vessel segmentation based on a low-resolution input. 2) We emphasize the interested features in two aspects, including an upsampling with a feature decomposition (UFD) module and a feature interaction module (FIM) with a new constraint loss. They extract the spatial correlation between the decomposed features and super-resolution features. 3) The efficacy of our proposed SuperVessel is shown on three publicly available datasets compared with other state-of-the-art algorithms.
\section{Methodology}
\begin{figure}
    \centering
    \includegraphics[width=1\textwidth]{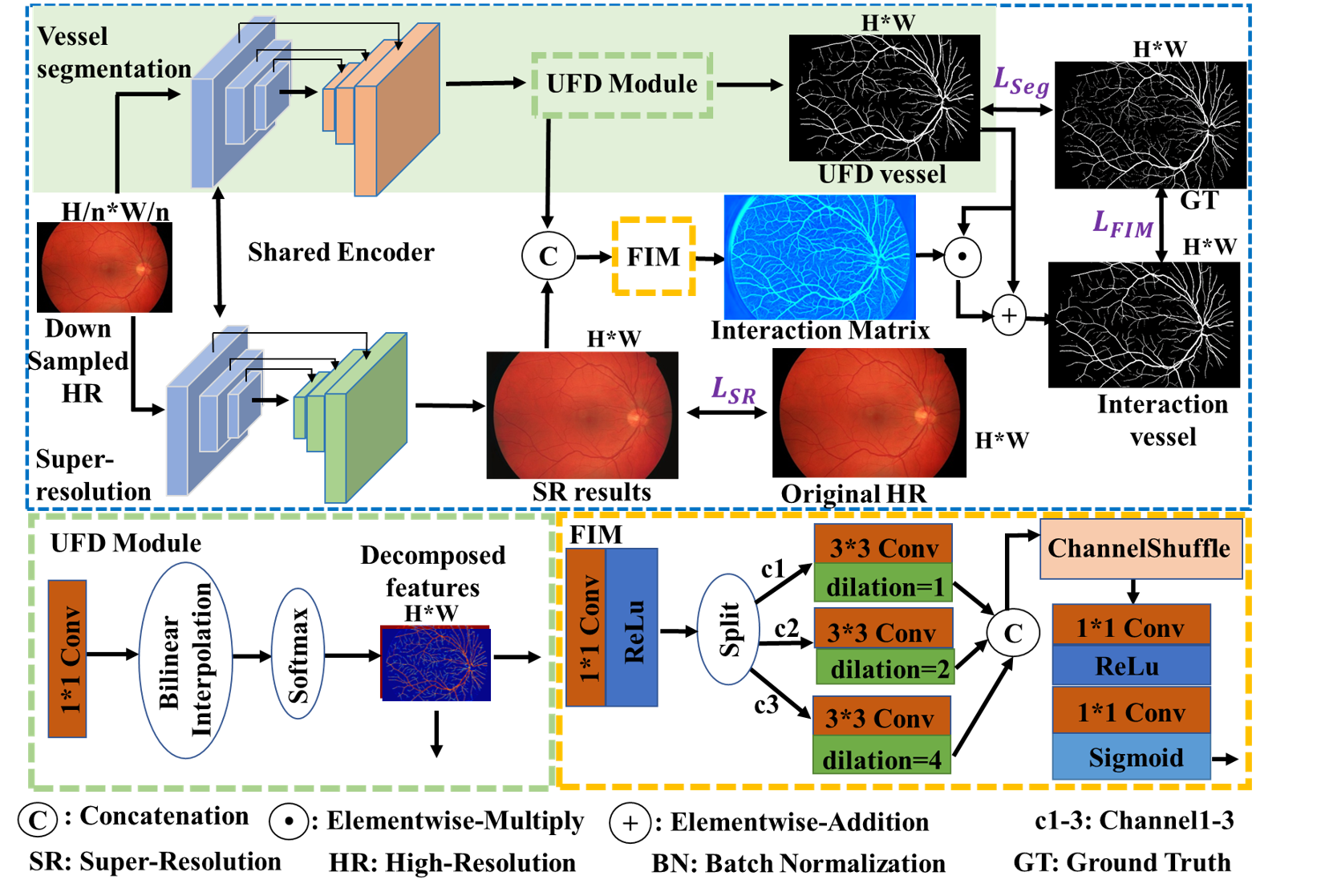}
    \caption{The pipeline of the proposed SuperVessel framework. For the test phase, only the vessel segmentation branch is adopted, as shown in the light green box.}
    \label{fig:pipeline}
\end{figure}

The pipeline of our proposed SuperVessel is illustrated in Fig. \ref{fig:pipeline}. Given a retinal image $X$ of size $H \times W$, we first downsample the image by $n\times$ to simulate a low-resolution image, which is adopted as the input of the whole framework. To reconstruct more appealing vessels, we propose two modules with a new loss function: An upsampling with feature decomposition (UFD) module separates vessel and background features into different channels. The proposed feature interaction module (FIM) emphasizes the vessel features by optimizing the feature interaction between the UFD module and the super-resolution branch. In the testing phase (as shown in the light green box), only the vessel segmentation branch is adopted to segment vessels to output high-resolution vessel segmentation results without extra computational load.
\subsection{SuperVessel Framework}
Decoded features only upsampled by bilinear interpolation cannot bring any additional information, since the input is a low-resolution image. Thus, we adopt super-resolution as an auxiliary network for vessel segmentation to provide more details in our SuperVessel framework, and the super-resolution network can be removed during the test phase. Ground truth for vessel segmentation is the labeled segmentation mask of the original high resolution, and that for super-resolution is the original high-resolution image. For the two branches, the same encoder-decoder structures \cite{ronneberger2015u} are adopted as the backbones. The encoder $E$ is shared, and two parallel decoders $D_{Seg}$ and $D_{SR}$ realize vessel segmentation and super-resolution, respectively, as shown in Fig. \ref{fig:pipeline}. Therefore, the whole structure of the SuperVessel could be formulated as :
\begin{gather}
    \mathrm{O_{Seg}} = \mathrm{UFD(D_{Seg}(E}(X))) \\
    \mathrm{O_{FIM}} = \mathrm{FIM(C(O_{Seg},O_{SR}))}\\ 
    \mathrm{O_{SR}} = \mathrm{D_{SR}(E}(X)) 
\end{gather}
where $\mathrm{O_{Seg}, O_{FIM}, and O_{SR}}$ are the output of vessel segmentation branch, the FIM module, and super-resolution branch, respectively. $X$ is the input image, $\mathrm{E}(X)$ is encoded features of image $X$, $\mathrm{D}$ is the corresponding decoder, $\mathrm{C}$ is the concatenation operation. 

The loss function of our framework is defined as: $\mathcal{L} = \mathcal{L}_{Seg} + \mathcal{L}_{SR} + \mathcal{L}_{FIM}$, where $\mathcal{L}_{Seg} = -\frac{1}{n}\sum_{i=0}^{n}GT_{i}\log{\mathrm{O_{Seg}}_{i}}$ for the loss between UFD vessel and GT,  $\mathcal{L}_{SR}(SR, HR) = \alpha * \left( SR - HR \right)^{2} +  (1-\alpha)* (1-\mathrm{SSIM}(SR, HR))$ for the loss of super-resolution branch, $\mathcal{L}_{FIM}$ for the loss between interaction vessel and GT, $n$ is the number of the classes, $GT$ is the label, $GT_i$ is the ground truth of the class $i$, and $\mathrm{O_{Seg}}^{i}$ is the probability of the class $i$ in the segmentation results. $SR$ is the predicted super-resolution image, $HR$ is the original high-resolution image as the ground truth. 

\subsection{Vessel Feature Enhancement}
To enhance the interested vessel features, we propose two modules, upsampling with feature decomposition (UFD) module and feature interaction module (FIM) with a loss $\mathcal{L}_{FIM}$. The former module splits the vessel features from the background, and the latter emphasizes the vessel features by capturing the spatial relation between segmentation and super-resolution branches.

\textbf{Upsampling with Feature Decomposition (UFD) Module: }
Previous algorithms constrain all features of the entire image by a loss function to the same degree. However, the background is not our interested target, and we hope the framework can focus on our interested vessels. Thus in our SuperVessel, we propose the upsampling with feature decomposition (UFD) module to split the background and vessel features, and the details are shown in the light green dotted frame of Fig. \ref{fig:pipeline}. The construction is simple but effective, only $1\times1$ Conv is adopted before bilinear interpolation \cite{press2007numerical} to output decomposed features in different channels. Then the features with two channels are input into our interaction module to obtain a vessel interaction matrix.

\textbf{Feature Interaction Module (FIM): } 
Most algorithms often fuse multiply tasks by various losses or similarities of entire images, which cannot focus on our interested vessel region, nor capture the spatial relation between separated segmentation features and super-resolution features.
As the structure information, like vessels, should correspond to the two branches, we propose a feature interaction module (FIM) to capture the spatial relation between features and mainly focus on our interested vessels. The detailed construction is shown in the yellow dotted frame of Fig. \ref{fig:pipeline}. The decomposed background, vessel features from segmentation, and super-resolution features are concatenated together into the FIM. $1\times1$ Conv with ReLU as the activation is adopted to map the input features into tensors with dimension $d$. The tensors are split into three groups based on channel, and dimension of each group is $d/3$. Then three $3\times3$ Conv with different dilation rates $dilation=1,2,4$ to capture different scale information from three groups. In this way, features with different scales can be obtained. Then we concatenate these features to be one tensor. One $1\times1$ Conv can be used to integrate information from different scales effectively, thus the spatial relevance can be obtained. To further emphasize the interaction between each group, we adopt ChannelShuffle \cite{zhang2018shufflenet} to exchange the information of concatenated features, which are output from three different dilated rates. Finally, $1 \times 1$ Conv with ReLU followed by one $1\times1$ Conv with the Sigmoid is adopted as the activation function to generate the weight matrix of spatial interaction.

The product of mask and high-resolution image is often adopted to produce the region of interest, which takes the whole image as an entire. This often brings some false similarity expressions, especially when some vessels labeled in the mask cannot clearly show up in the corresponding high-resolution image (such as blurry or hard to see). To solve such a problem, we propose to use the prediction of segmentation adding the product of the interaction matrix and prediction of segmentation, and take the segmentation mask of high-resolution as the ground truth. In this way, the framework focuses on the shared region of vessel structures. Thus the loss of FIM $\mathcal{L}_{FIM}$ is expressed as:
\begin{equation}
    \mathcal{L}_{FIM} = -\frac{1}{n}\sum_{i=0}^{n}GT_{i}\log{(\mathrm{O_{Seg} \odot O_{FIM}+O_{Seg}})_{i}}
\end{equation}
where $n$ is the number of the classes, $GT_i$ is the ground truth of the class $i$, $\mathrm{O_{Seg}}$ is the output of the segmentation, $\mathrm{O_{FIM}}$ is the output of the FIM, and $(\mathrm{O_{Seg} \odot O_{FIM}+O_{Seg})}_{i}$ is the probability of the class $i$.

\section{Experiments}
The vessel segment branch of our SuperVessel is adopted to conduct the following experiments. In the section, we first introduce the datasets, evaluation metrics, and experiment parameters. Then the ablation study is listed. Finally, the performance of our SuperVessel is evaluated compared with other state-of-the-art methods. 

\noindent\textbf{Datasets:} We evaluated our SuperVessel with three modals of retinal images from three publicly available datasets, including Color fundus (\textbf{HRF}) \cite{budai2013robust}, OCTA (\textbf{OCTA-6M}) \cite{li2020image}, and ultra-widefield retinal images (\textbf{PRIME-FP20}) \cite{ding2020weakly}:

\noindent\textbf{HRF:} The dataset contains 45 color fundus images from healthy person and patients with diabetic retinopathy or glaucoma. The image size is $3504 \times 2336$. 30  images are used for training, and the rest 15 images for test.

\noindent\textbf{OCTA-6M:} The dataset contains 300 subjects' images, from the OCTA-500 \cite{li2020image}. OCTA (Optical Coherence Tomography Angiography) is a novel non-invasive imaging modality that visualizes human retinal vascular details. The field of view is $6mm\times6mm$, with resolution $400\times 400$ pixels. We use the first 240 subjects to train the model, and the other 60 subjects for test. 

\noindent\textbf{PRIME-FP20:} The dataset provides 15 high-resolution ultra-widefield (UWF) fundus photography (FP) images using Optos 200Tx camera. All images have the same resolution $4000\times 4000$ pixels. \textcolor{black}{The first 10 images are used for training, and the rest for test}.

\noindent\textbf{Evaluation metrics:} The evaluation metrics include Precision(P), Sensitivity (SE), Intersection over Union (IoU), Dice, Accuracy(ACC), and Area under the ROC curve (AUC).






\noindent\textbf{Implementation Details:} All the experiments are run on one NVIDIA RTX 2080TI GPU. We used SGD as the optimizer with the momentum of 0.9 and the weight decay of 0.0001. We used the poly learning rate adjust schedule strategy \cite{liu2015parsenet} to set the learning rate during training, where $lr=((1-\frac{iter}{max\_iter})^{power})*init\_lr$, and we set $init\_lr=0.01, power=0.9$. In addition, the training epoch is set as $128$. Due to the memory limit, we cannot use the original size of the HRF and PRIME-FP datasets to train the model, we use $1752\times 1162$ and $1408\times 1296$ as the target size of the high-resolution image for these two datasets respectively. 

\subsection{Ablation Study}
An ablation study is conducted on the HRF dataset to investigate the effectiveness of designed modules in our SuperVessel, which takes U-net as the backbone. We also counted the computational parameters and FLOPs of our SuperVessel in both training and test phases. The parameters of SuperVessel for training and test are 29.73M and 28.95M, respectively. Its FLOPs for training and test are 5.86G and 4.72G, respectively. The parameters and FLOPs of SuperVessel for the test are the same as those of U-net. Thus our SuperVessel does not increase the computation load.
\begin{table}[htbp]
    \centering
    \caption{Ablation Study of SuperVessel $\left(mean\pm \mathrm{std}\right)$. }
    \resizebox{0.9\textwidth}{!}{
    \begin{tabular}{ccc cccccc}
        \toprule
        ASR & UFD & FIM & SE & IoU & Dice & ACC & AUC
        \\
        \midrule
         \text{\sffamily x} &  \text{\sffamily x} &  \text{\sffamily x}&$68.56\pm1.96$&$56.73\pm0.94$&$72.39\pm0.76$&$95.87\pm0.04$&$83.05\pm0.96$
         \\
         \checkmark & \text{\sffamily x} & \text{\sffamily x} &$68.40\pm2.38$&$56.13\pm1.01$&$71.89\pm0.83$&$95.78\pm0.10$&$82.89\pm1.08$
         \\
         \checkmark & \checkmark  &  \text{\sffamily x}&$68.31\pm2.13$&$59.41\pm0.97$&$74.53\pm0.77$&$96.32\pm0.05$&$83.08\pm1.05$
         \\
         \checkmark & \checkmark & \checkmark&$\bold{72.29}\pm1.30$&$\bold{62.26}\pm0.38$&$\bold{76.74}\pm0.29$&$\bold{96.54}\pm0.03$&$\bold{85.06}\pm0.66$
         \\
         \bottomrule
    \end{tabular}
    }
    \label{tab:ablation_study}
\end{table}

As shown in Table \ref{tab:ablation_study}, we proposed three modules, including ASR (auxiliary super-resolution task), UFD, and FIM. Only adding ASR, meaning purely adding a super-resolution branch to the vessel segmentation, IoU and Dice decrease a little, so simply combining the two tasks does not effectively improve the segmentation results. After UFD is inserted into the network, the IoU and Dice further increase by about 3\% than the baseline, which illustrates that separating vessel features from the background can make the network focus on the vessel features. Finally, the network with FIM improves the IoU and Dice by about 6\% and 4\%, which means that the interaction between the two branches can further emphasize the vessel features. Our feature enhancement can effectively improve the segmentation accuracy for our SuperVessel. Therefore, the SuperVessel improves the vessel segmentation accuracy without increasing the computation load.

\subsection{Comparison Results}
Seven state-of-the-art methods are selected for comparison, four vessel segmentation networks including U-net \cite{ronneberger2015u}, SA-UNet \cite{guo2021sa}, CS-Net \cite{10.1007/978-3-030-32239-7_80}, SCS-Net \cite{WU2021102025}; three super-resolution-combined segmentation multi-task networks including DSRL \cite{Wang_2020_CVPR}, CogSeg \cite{9506999} and PFSeg \cite{wang2021patch}. The proposed SuperVessel is compared with the above methods based on the three vessel segmentation datasets.  

\begin{table}[htbp]
    \centering
    \caption{Results on the HRF dataset ($\mathrm{mean}\pm \mathrm{std})$. }
    \resizebox{0.9\textwidth}{!}{
        \begin{tabular}{ccccccccc}
             \toprule 
            & Model &   SE & IoU & Dice & ACC & AUC \\
             \midrule
             &U-net &$68.56\pm1.96$&$56.73\pm0.94$&$72.39\pm0.76$&$95.87\pm0.04$&$83.05\pm0.96$
             \\
            &SA-UNet&$68.19\pm2.89$&$55.26\pm1.02$&$71.18\pm0.84$&$95.64\pm0.07$&$82.70\pm1.33$
             \\
           &CS-Net&$67.43\pm2.45$&$55.09\pm0.92$&$71.04\pm0.76$&$95.66\pm0.08$&$82.32\pm1.13$
             \\
              &SCS-Net&$66.31\pm1.76$&$54.63\pm0.30$&$70.66\pm0.25$&$95.65\pm0.13$&$81.80\pm0.73$
             \\ 
             \cline{2-7}
             &DSRL&-&-&-&-&-
             \\
            &CogSeg&$70.95\pm1.72$&$59.68\pm0.76$&$74.75\pm0.6$&$96.22\pm0.04$&$84.31\pm0.81$
             \\
             &PFSeg&-&-&-&-&-
             \\
             &\textbf{SuperVessel}&$\bold{72.29}\pm1.30$&$\bold{62.26}\pm0.38$&$\bold{76.74}\pm0.29$&$\bold{96.54}\pm0.03$&$\bold{85.06}\pm0.66$
             \\
            \bottomrule
        \end{tabular}
    }
    \label{tab:exphrf}
\end{table}

\noindent\textbf{Comparison Results on HRF datasets:} The sizes of input images for all the algorithms are the same $876\times584$, to simulate low-resolution input images. The sizes of output images for U-net, SA-UNet, CS-Net and SCS-Net are the same as their input, and those for DSRL, CogSeg, PFSeg and our SuperVessel are $1752\times 1162$. From Table \ref{tab:exphrf}, our SuperVessel surpasses all the other state-of-the-art networks, with an IoU of more than 2\%. The std numbers of our SuperVessel are the lowest, meaning that the segmentation stability of SuperVessel is superior to other algorithms. DSRL and PFSeg cannot segment the HRF dataset, as the gradient explosion happens during training. We will discuss this in the discussion section. Thus, the superiority of our SuperVessel in the accuracy and stability can be proved on the HRF dataset.
\begin{figure}
    \centering
    \includegraphics[width=1\textwidth]{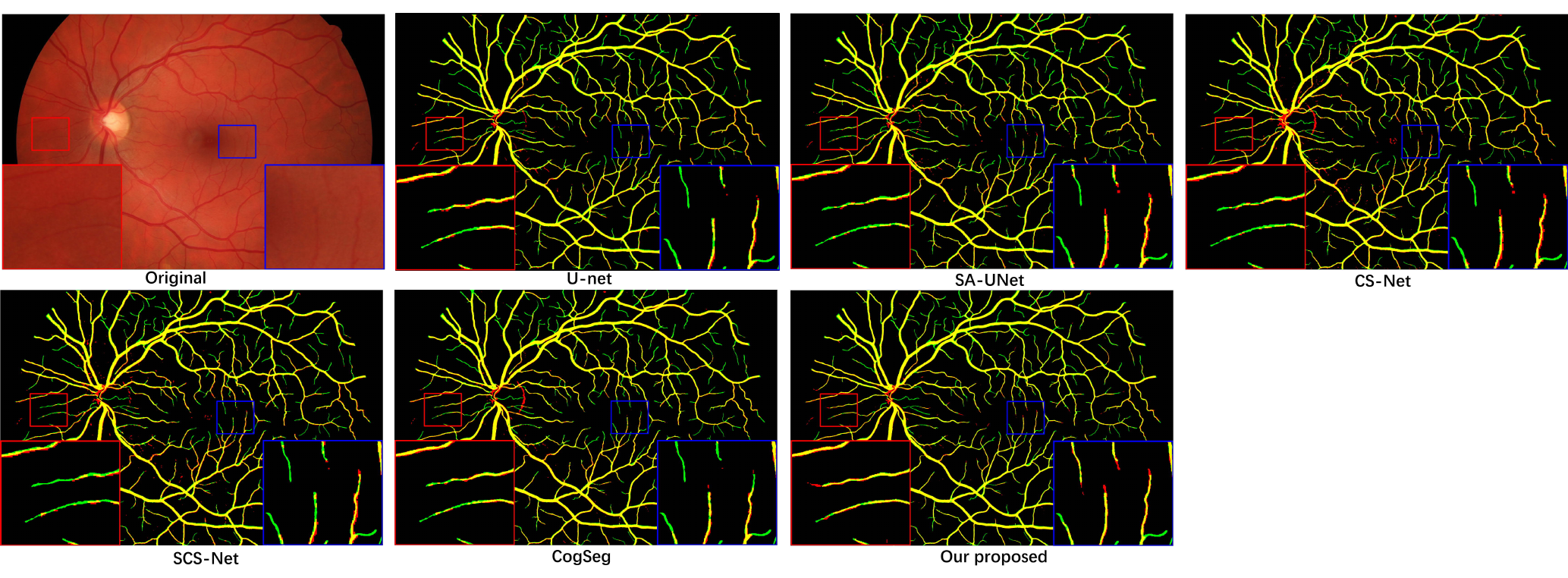}
    \caption{The experiment examples on the HRF datasets. Green means the ground truth, red is segmentation result, yellow means the corrected segmentation results. \textbf{(Please zoom in for a better view.)}}
    \label{fig:exphrf}
\end{figure}

The experiment examples on HRF dataset are shown in Fig. \ref{fig:exphrf}. All the other algorithms wrongly segment the edge of the disc as vessels, but our SuperVessel gives out the exact classification. Then we selected two blocks with tiny vessels to further analyze the results, the blue rectangle contains tiny vessels around the macular, and the red rectangle contains the vessels' end. The SuperVessel can segment more tiny vessels than other algorithms, especially, at the end of all vessels. Thus, the proposed feature enhancement can effectively improve the tiny vessel segmentation.
\begin{table}[htbp]
    \centering
        \caption{Comparison results based on OCTA dataset.}
        \resizebox{0.9\textwidth}{!}{
        \begin{tabular}{ccccccccc}
             \toprule 
            &Model &   SE & IoU & Dice & ACC & AUC \\
            \midrule 
             &U-net &$64.06\pm0.41$&$52.36\pm0.18$&$68.73\pm0.15$&$94.53\pm0.01$&$80.97\pm0.18$
             \\
           &SA-UNet&$59.65\pm0.41$&$48.94\pm0.25$&$65.72\pm0.23$&$94.16\pm0.02$&$78.86\pm0.20$
             \\
             &CS-Net&$61.58\pm0.84$&$50.32\pm0.25$&$66.95\pm0.22$&$94.29\pm0.03$&$79.79\pm0.36$
             \\
             &SCS-Net&$63.44\pm0.27$&$51.17\pm0.04$&$67.70\pm0.04$&$94.32\pm0.02$&$80.62\pm0.12$
             \\ 
             \cline{2-7}
            &DSRL&-&-&-&-&-
             \\
             &CogSeg&$66.33\pm0.27$&$56.46\pm0.48$&$72.17\pm0.39$&$95.2\pm0.08$&$82.42\pm0.16$
             \\
             &PFSeg&$\bold{74.92}\pm2.68$&$56.35\pm2.55$&$72.05\pm2.07$&$94.55\pm0.39$&$86.02\pm1.3$
             \\
             &\textbf{SuperVessel}&$73.80\pm0.31$&$\bold{64.56}\pm0.10$&$\bold{78.46}\pm0.07$&$\bold{96.20}\pm0.02$&$\bold{86.30}\pm0.13$
             \\
             \bottomrule
    \end{tabular}
    }
    \label{tab:expocta}
\end{table}

\noindent\textbf{Comparison Results on OCTA dataset: }
The sizes of low-resolution input images for all the algorithms are the same $200\times 200$. The sizes of output images for U-net, SA-UNet, CS-Net and SCS-Net are the same as their input, and those for DSRL, CogSeg, PFSeg and our SuperVessel are $400\times 400$. From Table \ref{tab:expocta}, our SuperVessel surpasses all the other state-of-the-art networks, with an IoU of more than 8\%. The std numbers of our SuperVessel are lower than most other algorithms, meaning that the SuperVessel works stably. We will also discuss that DSRL cannot segment the OCTA dataset in the discussion section. 

\begin{figure}
    \centering
    \includegraphics[width=1\textwidth]{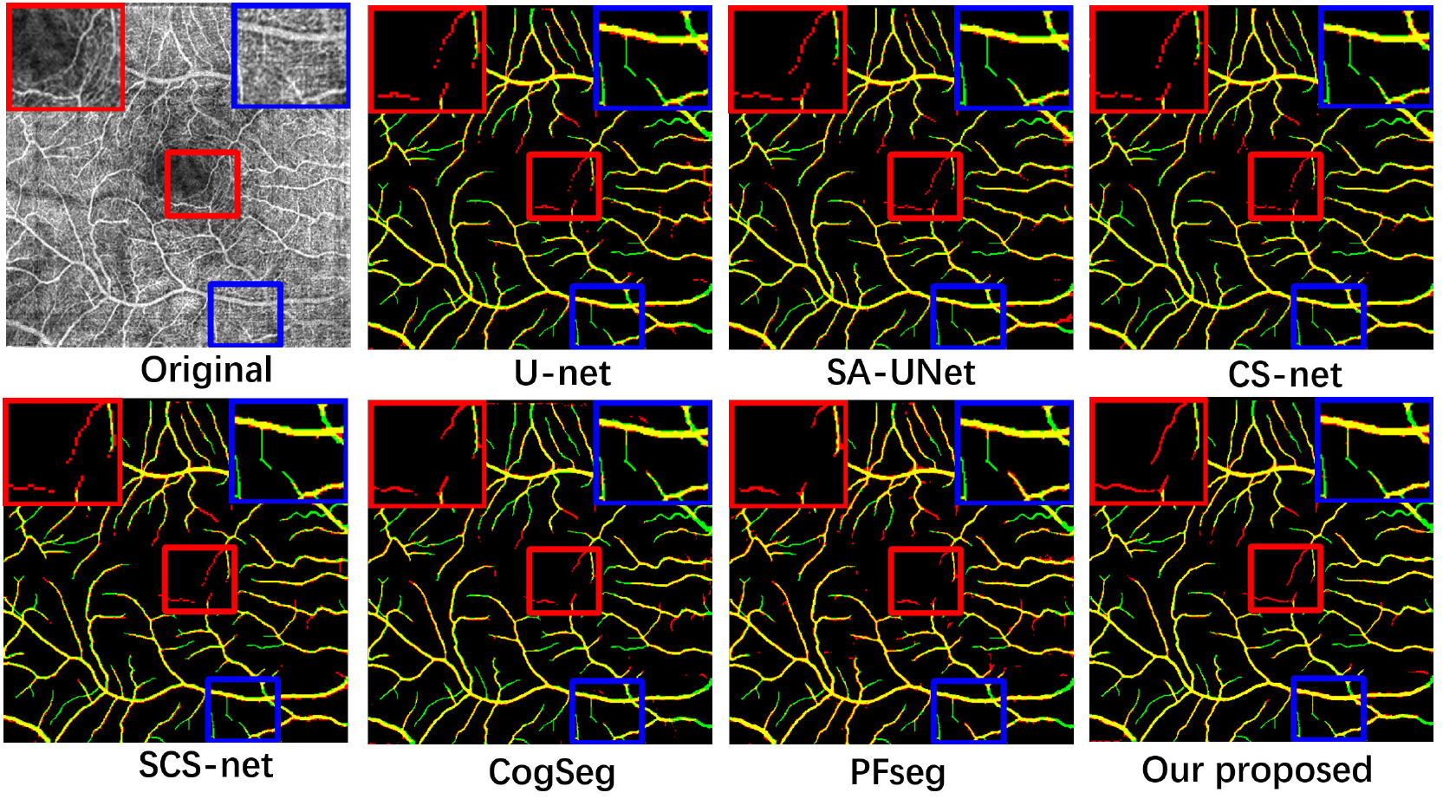}
    \caption{The experiment examples on the OCTA datasets. Green means the ground truth, red is segmentation result, yellow means the corrected segmentation results. \textbf{(Please zoom in for a better view.)}}
    \label{fig:expocta}
\end{figure}

The experiment examples on the OCTA dataset are shown in Fig. \ref{fig:expocta}. Most comparison algorithms produce some discontinue vessels as shown in the red rectangles, as the vessels around the macular are very tiny and indistinguishable. There are two tiny vessels away from the large vessels in the blue rectangles. Our SuperVessel can detect the tiny vessels, but other algorithms cannot correctly segment them, since our SuperVessel highlights the structure features based on the enhancement of the features. Thus, the superiority of our SuperVessel in the accuracy, segmenting of tiny vessels, and stability can be proved on the OCTA dataset.

\begin{table}[htbp]
    \centering
        \caption{Comparison results on PRIME-FP20 dataset}
         \resizebox{0.9\textwidth}{!}{
         \begin{tabular}{ccccccccc}
             \toprule 
            &Model &   SE & IoU & Dice & ACC & AUC \\
            \midrule
             &U-net&$26.43\pm1.34$&$22.77\pm0.89$&$37.08\pm1.18$&$97.77\pm0.01$&$62.62\pm0.70$ 
            \\
            &SA-UNet&$13.50\pm5.90$&$12.43\pm5.33$&$21.77\pm8.94$&$97.66\pm0.08$&$56.47\pm2.82$
            \\
             &CS-Net&$19.37\pm6.90$&$17.06\pm5.31$&$28.86\pm7.91$&$97.70\pm0.04$&$59.26\pm3.24$
            \\ 
            &SCS-Net&$18.91\pm4.68$&$17.08\pm3.72$&$29.04\pm5.40$&$97.73\pm0.04$&$59.05\pm2.19$
            \\ 
            \cline{2-7}
            &DSRL&-&-&-&-&-
            \\
          &CogSeg&$11.14\pm7.79$&$10.50\pm7.25$&$18.36\pm12.31$&$97.69\pm0.13$&$55.52\pm3.82$
            \\
            & PFSeg&-&-&-&-&-
            \\
            &\textbf{SuperVessel}&$\bold{38.47}\pm1.7$&$\bold{33.52}\pm1.12$&$\bold{50.21}\pm1.25$&$\bold{98.11}\pm0.03$&$\bold{68.67}\pm0.84$
            \\
            \bottomrule
    \end{tabular}
    }
    \label{tab:expprime}
\end{table}

\begin{figure}
    \centering
    \includegraphics[width=1\textwidth]{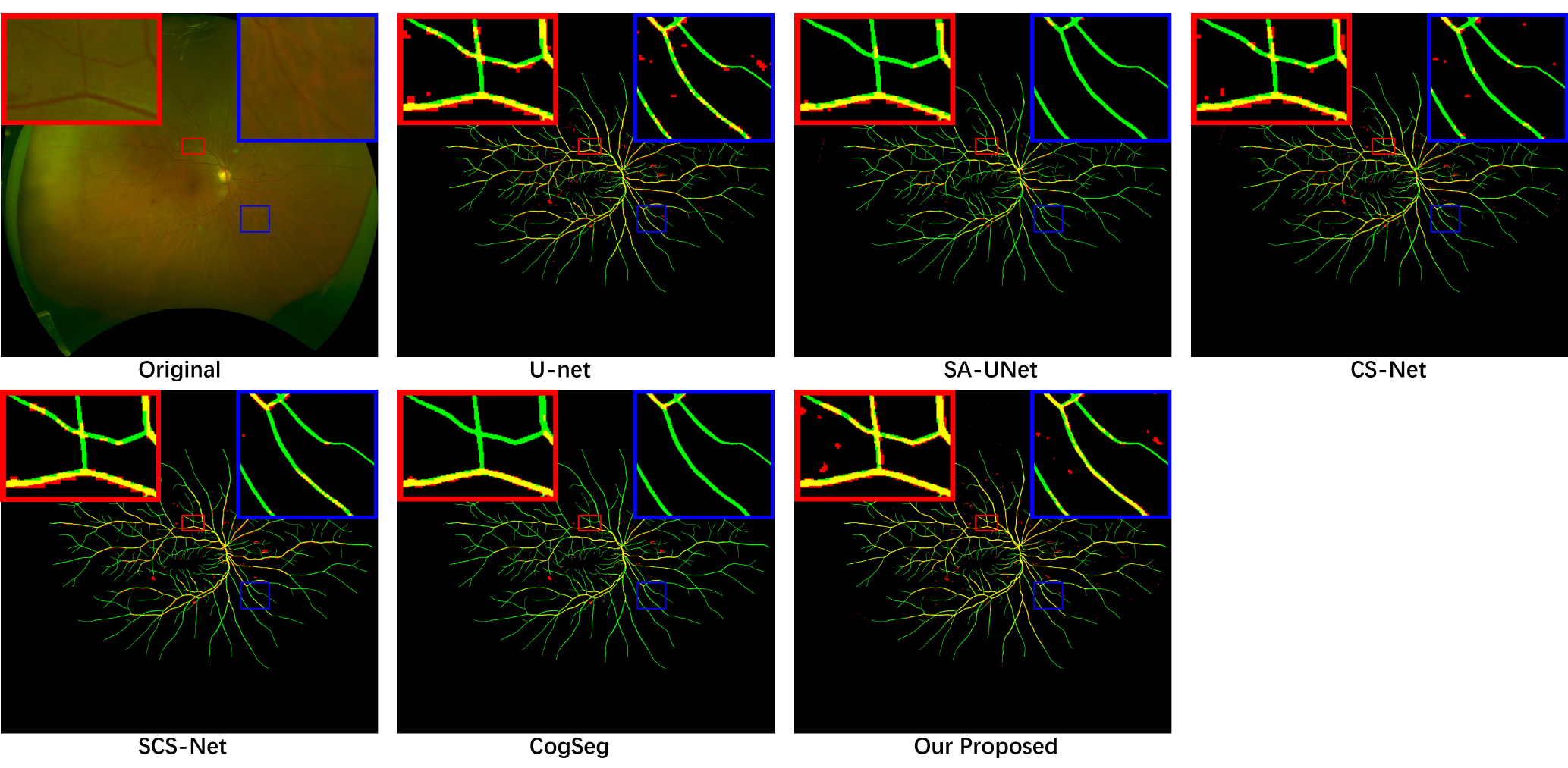}
    \caption{The experiment examples on the PRIME-FP20 dataset. Green means the ground truth, red is segmentation result, yellow means the corrected segmentation results. \textbf{(Please zoom in for a better view.)}}
    \label{fig:expprime}
\end{figure}

\noindent\textbf{Comparison results on PRIME-FP20 dataset: }
The sizes of input images for all the algorithms are the same $704\times 648$, to simulate low-resolution input images. The sizes of output images for U-net, SA-UNet, CS-Net and SCS-Net are the same as their input, and those for DSRL, CogSeg, PFSeg and our SuperVessel are $1408\times 1296$. From Table \ref{tab:expprime}, our SuperVessel surpasses all the other state-of-the-art networks, with IoU of more than 11\%, which is significant. The std parameters of our SuperVessel are lower than most of the other algorithms, meaning that the SuperVessel works stably. For the situation of DSRL and PFSeg, we will also discuss this in the discussion section. 
The experiment examples on the dataset are shown in Fig. \ref{fig:expprime}. As the view field of these images is about $200^{\circ}$, the vessels in these images are extremely tiny compared with the other two datasets. In the red rectangles, the segmented vessels by our SuperVessel are more continuous than those by other algorithms. In the blue rectangles containing more tiny vessels, the SuperVessel detects more vessels than other algorithms, as the spatial features such as vessels are emphasized by our feature enhancement. Therefore, our SuperVessel is superior to other state-of-the-art algorithms in the segmentation accuracy and stability with tiny vessels based on the three publicly available datasets. 
\section{Discussion and Conclusions}
In the study, we proposed the SuperVessel to provide high-resolution vessel segmentation results for analysis based on low-resolution input, and experiments prove its effectiveness. But previous super-resolution-combined segmentation multi-task networks such as DSRL and PFSeg cannot train on most of the vessel segmentation datasets, we observed that these methods often cause the model to collapse during training. We conjectured that the similarity loss between two tasks is not suitable when the targets of one task are the subset of another task. Since the similarity maybe make one of the tasks lose its constraint direction, the collapsing of models maybe happen, such as the two algorithms collapsed in the vessel segmentation task. Experiments on three datasets show that our proposed method can resolve this problem by finding the spatial relation between two tasks for vessel segmentation. 

Although the SuperVessel can work well on most of the vessel segmentation datasets, there are still some improvements. For some extremely tiny blood vessels, especially for ultra-field fundus images in the PRIME-FP20 dataset, the model can only segment a little more vessels than other algorithms. The main reason is that the original image is very large but we cannot deal with so much redundant information. Limited by our computation devices we cannot train our SuperVessel to output the vessel of original image size for some extremely large-image-size datasets. 

In conclusion, we proposed the SuperVessel for vessel segmentation, which outputs high-resolution vessel segmentation results based on low-resolution input images. Experiments on three publicly available datasets prove that super-resolution branches could provide detailed features for vessel segmentation, and the proposed feature enhancement, which focuses on target features, can further improve the segmentation accuracy with more tiny vessels and stronger continuity of the segmented vessels. 

\section{Acknowledgement}
This work was supported in part by Guangdong Provincial Department of Education (2020ZDZX3043), Guangdong Provincial Key Laboratory (2020B121201001), and Shenzhen Natural Science Fund (JCYJ20200109140820699 and the Stable Support Plan Program 20200925174052004).
\bibliographystyle{splncs04}
\bibliography{ref}

\begin{thebibliography}{10}
\providecommand{\url}[1]{\texttt{#1}}
\providecommand{\urlprefix}{URL }
\providecommand{\doi}[1]{https://doi.org/#1}

\bibitem{budai2013robust}
Budai, A., Bock, R., Maier, A., Hornegger, J., Michelson, G.: Robust vessel
  segmentation in fundus images. International journal of biomedical imaging
  \textbf{2013} (2013)

\bibitem{ding2020weakly}
Ding, L., Kuriyan, A.E., Ramchandran, R.S., Wykoff, C.C., Sharma, G.:
  Weakly-supervised vessel detection in ultra-widefield fundus photography via
  iterative multi-modal registration and learning. IEEE Transactions on Medical
  Imaging  (2020)

\bibitem{fu2016deepvessel}
Fu, H., Xu, Y., Lin, S., Wong, D.W.K., Liu, J.: Deepvessel: Retinal vessel
  segmentation via deep learning and conditional random field. In:
  International conference on medical image computing and computer-assisted
  intervention. pp. 132--139. Springer (2016)

\bibitem{guo2021sa}
Guo, C., Szemenyei, M., Yi, Y., Wang, W., Chen, B., Fan, C.: Sa-unet: Spatial
  attention u-net for retinal vessel segmentation. In: 2020 25th International
  Conference on Pattern Recognition (ICPR). pp. 1236--1242. IEEE (2021)

\bibitem{ikram2013retinal}
Ikram, M.K., Ong, Y.T., Cheung, C.Y., Wong, T.Y.: Retinal vascular caliber
  measurements: clinical significance, current knowledge and future
  perspectives. Ophthalmologica  \textbf{229}(3),  125--136 (2013)

\bibitem{karaali2021dr}
Karaali, A., Dahyot, R., Sexton, D.J.: Dr-vnet: Retinal vessel segmentation via
  dense residual unet. arXiv preprint arXiv:2111.04739  (2021)

\bibitem{DBLP:conf/igarss/LeiSWPXH19}
Lei, S., Shi, Z., Wu, X., Pan, B., Xu, X., Hao, H.: Simultaneous
  super-resolution and segmentation for remote sensing images. In: 2019 {IEEE}
  International Geoscience and Remote Sensing Symposium, {IGARSS} 2019,
  Yokohama, Japan, July 28 - August 2, 2019. pp. 3121--3124. {IEEE} (2019)

\bibitem{DBLP:conf/wacv/LiVNNK20}
Li, L., Verma, M., Nakashima, Y., Nagahara, H., Kawasaki, R.: Iternet: Retinal
  image segmentation utilizing structural redundancy in vessel networks. In:
  {IEEE} Winter Conference on Applications of Computer Vision, {WACV} 2020,
  Snowmass Village, CO, USA, March 1-5, 2020. pp. 3645--3654. {IEEE} (2020)

\bibitem{li2020image}
Li, M., Chen, Y., Ji, Z., Xie, K., Yuan, S., Chen, Q., Li, S.: Image projection
  network: 3d to 2d image segmentation in octa images. IEEE Transactions on
  Medical Imaging  \textbf{39}(11),  3343--3354 (2020)

\bibitem{liu2015parsenet}
Liu, W., Rabinovich, A., Berg, A.C.: Parsenet: Looking wider to see better.
  arXiv preprint arXiv:1506.04579  (2015)

\bibitem{london2013retina}
London, A., Benhar, I., Schwartz, M.: The retina as a window to the
  brain—from eye research to cns disorders. Nature Reviews Neurology
  \textbf{9}(1),  44--53 (2013)

\bibitem{10.1007/978-3-030-32239-7_80}
Mou, L., Zhao, Y., Chen, L., Cheng, J., Gu, Z., Hao, H., Qi, H., Zheng, Y.,
  Frangi, A., Liu, J.: Cs-net: Channel and spatial attention network for
  curvilinear structure segmentation. In: Shen, D., Liu, T., Peters, T.M.,
  Staib, L.H., Essert, C., Zhou, S., Yap, P.T., Khan, A. (eds.) Medical Image
  Computing and Computer Assisted Intervention -- MICCAI 2019. pp. 721--730.
  Springer International Publishing, Cham (2019)

\bibitem{press2007numerical}
Press, W.H., Teukolsky, S.A., Vetterling, W.T., Flannery, B.P.: Numerical
  recipes 3rd edition: The art of scientific computing. Cambridge university
  press (2007)

\bibitem{ronneberger2015u}
Ronneberger, O., Fischer, P., Brox, T.: U-net: Convolutional networks for
  biomedical image segmentation. In: International Conference on Medical image
  computing and computer-assisted intervention. pp. 234--241. Springer (2015)

\bibitem{su2021msu}
Su, R., Zhang, D., Liu, J., Cheng, C.: Msu-net: Multi-scale u-net for 2d
  medical image segmentation. Frontiers in Genetics  \textbf{12}, ~140 (2021)

\bibitem{sun2009retinal}
Sun, C., Wang, J.J., Mackey, D.A., Wong, T.Y.: Retinal vascular caliber:
  systemic, environmental, and genetic associations. Survey of ophthalmology
  \textbf{54}(1),  74--95 (2009)

\bibitem{DBLP:conf/miccai/Wang0H19}
Wang, B., Qiu, S., He, H.: Dual encoding u-net for retinal vessel segmentation.
  In: Shen, D., Liu, T., Peters, T.M., Staib, L.H., Essert, C., Zhou, S., Yap,
  P., Khan, A.R. (eds.) Medical Image Computing and Computer Assisted
  Intervention - {MICCAI} 2019 - 22nd International Conference, Shenzhen,
  China, October 13-17, 2019, Proceedings, Part {I}. Lecture Notes in Computer
  Science, vol. 11764, pp. 84--92. Springer (2019)

\bibitem{wang2021patch}
Wang, H., Lin, L., Hu, H., Chen, Q., Li, Y., Iwamoto, Y., Han, X.H., Chen,
  Y.W., Tong, R.: Patch-free 3d medical image segmentation driven by
  super-resolution technique and self-supervised guidance. In: International
  Conference on Medical Image Computing and Computer-Assisted Intervention. pp.
  131--141. Springer (2021)

\bibitem{Wang_2020_CVPR}
Wang, L., Li, D., Zhu, Y., Tian, L., Shan, Y.: Dual super-resolution learning
  for semantic segmentation. In: Proceedings of the IEEE/CVF Conference on
  Computer Vision and Pattern Recognition (CVPR) (June 2020)

\bibitem{WU2021102025}
Wu, H., Wang, W., Zhong, J., Lei, B., Wen, Z., Qin, J.: Scs-net: A scale and
  context sensitive network for retinal vessel segmentation. Medical Image
  Analysis  \textbf{70},  102025 (2021)

\bibitem{9506999}
Zhang, Q., Yang, G., Zhang, G.: Collaborative network for super-resolution and
  semantic segmentation of remote sensing images. IEEE Transactions on
  Geoscience and Remote Sensing  \textbf{60},  1--12 (2022)

\bibitem{zhang2018shufflenet}
Zhang, X., Zhou, X., Lin, M., Sun, J.: Shufflenet: An extremely efficient
  convolutional neural network for mobile devices. In: Proceedings of the IEEE
  conference on computer vision and pattern recognition. pp. 6848--6856 (2018)

\end{thebibliography}
%





\end{document}